\documentclass[twocolumn,english,prb,superscriptaddress,longbibliography]{revtex4-2}
\usepackage{color}
\usepackage[utf8]{inputenc}
\usepackage{graphicx}
\usepackage[T1]{fontenc}
\usepackage{float}
\usepackage{textcomp}
\usepackage{amsmath}
\usepackage{amssymb}
\usepackage{graphicx}
\usepackage{esint}
\usepackage{xcolor}
\usepackage{multirow}
\usepackage{ulem}
\usepackage{comment}
\graphicspath{{Figures/}}
\usepackage{dsfont}
\usepackage{soul}

\usepackage[colorlinks=true,hyperfootnotes=true,breaklinks=true]{hyperref}
\usepackage[capitalise]{cleveref}

\begin{document}
\title{Negative charging energy in LaAlO$_3$/SrTiO$_3$ quantum dots induced by SrTiO$_3$ polarization}
 
\author{J. Czarnecki}
\affiliation{AGH University of Krakow, Faculty of Physics and
Applied Computer Science, al. Mickiewicza 30, 30-059 Kraków, Poland}

\author{A. Sierant}
\affiliation{AGH University of Krakow, Faculty of Physics and
Applied Computer Science, al. Mickiewicza 30, 30-059 Kraków, Poland}

\author{B. Szafran}
\affiliation{AGH University of Krakow, Faculty of Physics and
Applied Computer Science, al. Mickiewicza 30, 30-059 Kraków, Poland}

\author{P. Szumniak}
\affiliation{AGH University of Krakow, Faculty of Physics and
Applied Computer Science, al. Mickiewicza 30, 30-059 Kraków, Poland}

\author{T. S. Jespersen}
\affiliation{Department of Energy Conversion and Storage, Technical University of Denmark, Kgs. Lyngby, Denmark}

\author{G. Meucci}
\affiliation{Department of Energy Conversion and Storage, Technical University of Denmark, Kgs. Lyngby, Denmark}

\author{T. Jansen}
\affiliation{Department of Energy Conversion and Storage, Technical University of Denmark, Kgs. Lyngby, Denmark}

\author{R. Citro}
\affiliation{Physics Department ‘E. R. Caianiello’ and CNR-SPIN, Università degli Studi di Salerno, Fisciano, (SA), Italy}

\author{A. Maiellaro}
\affiliation{CNR-SPIN, c/o Università di Salerno, IT-84084 Fisciano (SA), Italy}
\affiliation{Physics Department ‘E. R. Caianiello’ and CNR-SPIN, Università degli Studi di Salerno, Fisciano, (SA), Italy}

\author{N. Bergeal}
\affiliation{Laboratoire de Physique et d'Etude des Mat\'eriaux, ESPCI Paris,
Universit\'e PSL, CNRS, Sorbonne Universit\'e, Paris, France}

\author{M. Bibes}
\affiliation{Laboratoire Albert Fert, CNRS, Thales, Universit´e Paris-Saclay, 91767 Palaiseau, France}

\author{M. Salluzzo}
\affiliation{CNR-SPIN, c/o Complesso di Monte S. Angelo, via Cinthia, 80126 Naples, Italy }

\author{A. Kalaboukhov}
\affiliation{Department of Microtechnology and Nanoscience (MC2), Chalmers University of Technology, 41296 Gothenburg, Sweden }

\author{P. Wójcik}
\email{pawelwojcik@agh.edu.pl}
\affiliation{AGH University of Krakow, Faculty of Physics and
Applied Computer Science, al. Mickiewicza 30, 30-059 Kraków, Poland}


\begin{abstract}
We propose a microscopic mechanism for the negative charging energy in LaAlO$_3$/SrTiO$_3$ quantum dots, arising from the electrostatic image charge associated with the paraelectric response of SrTiO$_3$. Using a self-consistent configuration-interaction approach that accounts for both Coulomb repulsion and image-charge interactions, we find a robust regime of electron pairing whose strength is tunable through the confinement potential and the effective image-charge distance, set by the thickness of the interfacial two-dimensional electron gas. Our results establish dielectric image-charge effects as a route to controllable electron pairing in oxide quantum dots and other materials with a strong dielectric response.
\end{abstract}

\maketitle


\textit{\textbf{Introduction -}} The electronic properties of the two-dimensional electron gas (2DEG) formed at interfaces between transition-metal oxides, such as LaAlO$_3$/SrTiO$_3$ (LAO/STO) and LaAlO$_3$/KTaO$_3$ (LAO/KTO), have attracted significant interest owing to their unique combination of high carrier mobility~\cite{Ohtomo2004}, strong spin-orbit (SO) coupling~\cite{Diez2015,Rout2017,Caviglia2010,Shalom2010,Yin2020}, gate-tunable superconductivity~\cite{Monteiro2017,Monteiro2019,joshua2012universal,jouan_multiband_sc_2022,Biscaras,Singh2017,Singh2019,Singh2019two,Trevisan2018,Wojcik_dome_schrodinger_poisson_sc_2024,Zegrodnik_2020,wojcik2021impact,Czarnecki}, magnetic ordering~\cite{Karen2012,Li2011,Brinkman2007,Bert2011}, and ferroelectricity~\cite{Gastiasoro2020,Kanasugi2018,Kanasugi2019,Honig2013}. The coexistence of these phenomena in a single material makes oxide interfaces an ideal platform for exploring electronic correlations and their evolution under reduced dimensionality, which in 2D oxides can be readily achieved through electrostatic confinement with a level of control comparable to that of semiconductor materials~\cite{Sumita2020,Trevisan2019,Pekker,Liu2023,Zhou_2025}. 

Electrostatically defined LAO/STO quantum dots (QDs) provide a particularly well-controlled platform for exploring electronic correlations in the few-electron regime~\cite{Guenevere2017,Cheng2015,Cheng2016}. Interestingly, the first experimental realizations of LAO/STO QDs~\cite{Jespersen2020,Guenevere2017,Cheng2015,Cheng2016} provide evidence for the negative charging energy, manifested by a preference for double occupancy of the QD even at magnetic fields and temperatures exceeding the critical values for superconductivity. Although this behavior has been described within the Anderson model with a negative-$U$ energy, the microscopic mechanism underlying this phenomenon remains unresolved. Understanding its origin may shed light not only on the electronic interactions in oxide-based QDs, but also on the still-unresolved mechanisms underlying electron pairing in oxide interfaces and other materials with a strong dielectric response.


\begin{figure}
    \centering
    \includegraphics[width=0.85\linewidth]{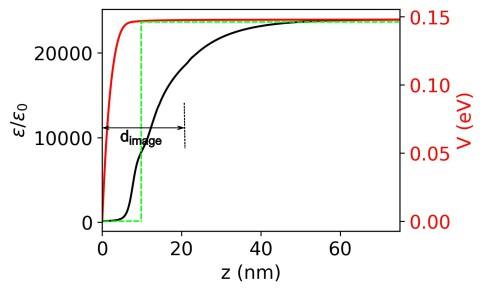}
    \put(-200,130){(a)}\\
    \includegraphics[width=0.85\linewidth]{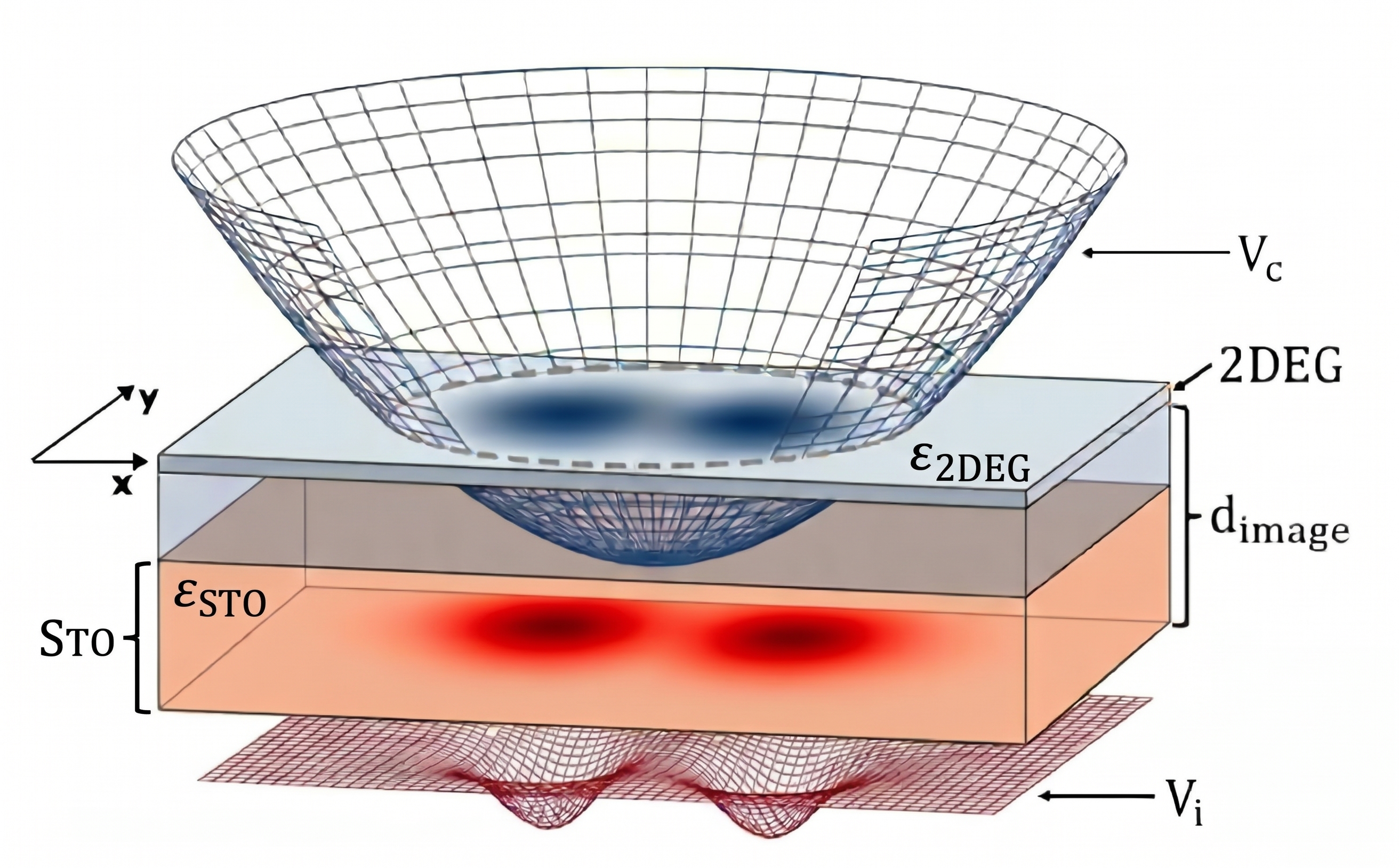}
    \put(-200,130){(b)}\\
    \caption{(a) Self-consistent potential $V(z)$ (red) and the dielectric permittivity $\varepsilon(z)$ (black) determined along the  direction perpendicular to the interface, located at $z=0$. The green line corresponds to the step-like profile approximation of $\varepsilon$ used in our calculations. It divides the system into two regions: (i) the 2DEG located at $z=0$ with $\varepsilon_{\mathrm{2DEG}} = 100$  and (ii) STO with $\varepsilon_{\mathrm{STO}} = 25000$, separated from the 2DEG by a distance $d_{\mathrm{image}}/2$.
    (b) Schematic illustration of the proposed mechanism. Electrons in the QD, confined by the external potential $V_{\mathrm{c}}$, induce image charge in STO, which generates an additional electrostatic potential $V_{\mathrm{i}}$ and leads to the negative charging energy. Parts of the schematic illustration was generated using Google Deep Mind's Nano Banana 2.}
    \label{fig1}
\end{figure}

In this work, we propose a microscopic mechanism for the negative charging energy (negative $U$) in LAO/STO QDs, which arises from their coupling to image charges induced by the paraelectric response of STO. The induced polarization is governed by the strong spatial variation of the dielectric permittivity near the LAO/STO interface~\cite{Mahabir2021}. The proposed mechanism is not limited to LAO/STO interfaces but may also apply to nanostructures embedded in other materials with a strong dielectric response.\\

\textit{\textbf{Mechanism of Negative Charging Energy -}}
The formation of the 2DEG at the LAO/STO interface is attributed to the strong electric field near the interface, which confines the electrons within an effective triangular quantum well~\cite{Popovic2008,Pavlenko2012}. The resulting electron gas has been reported to extend into STO over a characteristic length of approximately 5–15 nm~\cite{Sing2009,Salluzzo2009}. In Fig.~\ref{fig1}(a), we present the self-consistent potential profile $V(z)$ near the interface together with the corresponding electric permittivity $\varepsilon (z)$, obtained within the Schr\"odinger–Poisson method (see Ref.~\cite{Wojcik_dome_schrodinger_poisson_sc_2024} for details). In our calculations, we adopt the commonly used relation for $\varepsilon$ at $T=0$~K, given by
\begin{equation}
\varepsilon=\varepsilon_0+\frac{1}{A+B|F|},
\label{eq:eps}
\end{equation}
where $F$ is the electric field, $A=4.097\times 10^{-5}$, $B=4.907\times 10^{-10}$ m/eV and $\varepsilon_{0}=100$~\cite{jouan_multiband_sc_2022}.

As presented in Fig.~\ref{fig1}(a), the dielectric permittivity of STO varies strongly with position, taking relatively small values ($\varepsilon=100$) near the interface where the electric field is large and the 2DEG is created, and increases rapidly to values of the order of $25000$ over a distance of $30-40$~nm from the interface. Such a pronounced spatial variation of $\varepsilon$, together with the paraelectric nature of STO, gives rise to an effective image charge. When an electron is placed in the 2DEG, it polarizes STO. To determine the magnitude of the image charge analytically, we approximate the dielectric profile shown in Fig.~\ref{fig1}(a) (red line) by the step like function (green line) which divides the system into two distinct regions: (i) the 2DEG region at $z=0$, characterized by $\varepsilon_{2DEG} = 100$, and (ii) the bulk STO, located at a distance $d_{\mathrm{image}}/2$ from the 2DEG, with $\varepsilon_{\mathrm{STO}} = 25000$. Within this approximation, the image charge $q'$ has a sign opposite to that of the charge $q$ confined to the 2DEG, with its magnitude determined by the dielectric-permittivity mismatch across the interface (see the Supplemental Material)
\begin{equation}
    q'=-q\frac{\varepsilon_{STO}-\varepsilon_{2DEG}}{\varepsilon_{STO}+\varepsilon_{2DEG}}.
\end{equation}
For $\varepsilon_{\mathrm{STO}}=25000$ and $\varepsilon_{\mathrm{2DEG}}=100$, the image charge is close to $-q$, i.e., $q' \approx -q$.

The mechanism underlying the negative charging energy can be described as follows. When electrons are confined in the LAO/STO QD by an external confinement potential, they polarize STO, inducing an image charge of opposite sign and comparable magnitude. This image charge interacts electrostatically with the electrons in the QD. In particular, for a two-electron system, the image charge of one electron interacts not only with the electron itself but also attracts the other electron. For an appropriate value of $d_{\mathrm{image}}$, this interaction can lower the two-electron energy $E_t^{2e}$ below twice the single-electron energy $E_t^{1e}$, where the latter includes the interaction of the electron with its own image charge. The regime in which $E_t^{2e}-2E_t^{1e}<0$ constitutes a signature of the negative-$U$ regime, in which the quantum dot favors double occupancy. This condition corresponds to a negative difference between the chemical potentials of the two- and one-electron states of QD, $\Delta\mu=\mu_2-\mu_1$, where $\mu_N=E_t^N-E_t^{N-1}$, indicating that adding a second electron to the QD is energetically favorable (the negative charging energy).
This qualitative picture is presented in Fig.~\ref{fig1}(b) and confirmed below within a microscopic theoretical model that accounts for both Coulomb repulsion and image-charge interaction.\\

\textit{\textbf{Theory}} -
We consider two electrons in an LAO/STO QD modeled by a confinement potential, $V_c$. The single electron Hamiltonian is given by~\cite{Diez2015}
\begin{equation}
    \hat{H}_{\mathbf{k}}=\sum _{\mathbf{k}} \hat{C}^\dagger_{\mathbf{k}} ( \hat{H}_{0}+\hat{H}_{RSO}+\hat{H}_{SO} ) \hat{C}_{\mathbf{k}} + \sum_{i,j} V_{ij} \hat{C}^\dagger_{ij} \hat{C}_{ij},
    \label{eq:Hamiltonian_k_space}
\end{equation}
where $\hat{C}_{\mathbf{k}}=(\hat{c}_{\mathbf{k},xy}^{\uparrow}, \hat{c}_{\mathbf{k},xy}^{\downarrow}, \hat{c}_{\mathbf{k},xz}^{\uparrow}, \hat{c}_{\mathbf{k},xz}^{\downarrow}, \hat{c}_{\mathbf{k},yz}^{\uparrow}, \hat{c}_{\mathbf{k},yz}^{\downarrow})^{T}$ is the vector of annihilation operators for electrons with spin $\sigma=\uparrow,\downarrow$ on the orbital $d_{xy},d_{xz},d_{yz}$, in the state $\mathbf{k}$. The last term in the Hamiltonian corresponds to the potential energy defined in real space on the lattice, with site position $(i,j)$. It consists of two terms: (i) the confinement potential $V_c$, which defines the QD, and (ii) the potential arising from the image charge induced in STO, $V_i$, i.e. $V=V_c+V_i$. In Eq.~(\ref{eq:Hamiltonian_k_space}), $\hat{H}_0$ corresponds to the kinetic energy
and takes the form
\begin{equation}
\hat{H}_{0}=
\left(
\begin{array}{ccc}
 \epsilon^{xy}_{\mathbf{k}} & 0 & 0\\
 0 & \epsilon^{xz}_{\mathbf{k}} &  \epsilon^h_{\mathbf{k}} \\
 0 & \epsilon^h_{\mathbf{k}}  &  \epsilon^{yz}_{\mathbf{k}}
\end{array} \right) \otimes \hat {\sigma} _0\;,
\end{equation}
with  
\begin{equation}
\begin{split}
    \epsilon^{xy}_{\mathbf{k}}&=4t_l-2t_l\cos{k_x}-2t_l\cos{k_y}-\Delta_E,\\
    \epsilon^{xz}_{\mathbf{k}}&=2t_l+2t_h-2t_l\cos{k_x}-2t_h\cos{k_y},\\
    \epsilon^{yz}_{\mathbf{k}}&=2t_l+2t_h-2t_h\cos{k_x}-2t_l\cos{k_y},
\end{split}
\label{eq:H0}
\end{equation}
and the hybridization energy
\begin{equation}
\epsilon^h_{\mathbf{k}}=2t_d\sin{k_x}\sin{k_y}. 
\end{equation}
The remaining two terms in Eq.~(\ref{eq:Hamiltonian_k_space}), $\hat{H}_{SO}$ and $\hat{H}_{\mathrm{RSO}}$, describe the atomic $\mathbf{L} \cdot \mathbf{S}$ coupling and the Rashba SO interaction, respectively~\cite{Khalsa2013}. They can be written as
\begin{equation}
\hat{H}_{SO}= \frac{\Delta_{SO}}{3}
\left(
\begin{array}{ccc}
0 & i \sigma _x & -i \sigma _y\\
-i \sigma _x & 0 & i \sigma _z \\
i \sigma _y & -i \sigma _z & 0
\end{array} \right) \;,
\label{eq:hso}
\end{equation}
\begin{equation}
\hat{H}_{RSO}= \Delta_{RSO}
\left(
\begin{array}{ccc}
0 & i \sin{k_y} & i \sin{k_x}\\
-i \sin{k_y} & 0 & 0 \\
-i \sin{k_x} & 0 & 0
\end{array} \right) \otimes \hat {\sigma} _0\;,
\label{eq:rso}
\end{equation}
where $\Delta _{SO}$ and $\Delta _{RSO}$ determine the strength of the corresponding SO interactions, and $\sigma _{x},\sigma_{y},\sigma_{z}$ are the Pauli matrices. \\
The Hamiltonian of two electrons can be written as
\begin{equation}
    \hat{H}_2 = \hat{H}(1) + \hat{H}(2) + \frac{e^2}{4\pi\epsilon_0\epsilon_{2DEG} r_{12}}
    \label{eq:two_electrons}
\end{equation}
where $\hat{H}$ is the single-electron Hamiltonian given by Eq.~(\ref{eq:Hamiltonian_k_space}) and the last term represents the Coulomb interaction, with the strength determined by the dielectric constant $\epsilon_{2DEG}$. To treat the Coulomb interaction exactly, the two-electron problem is solved using the configuration interaction (CI) method in real space, as described in detail in our previous papers~\cite{Szafran_EDSR_2024,Szafran_manipulation_2024}. In the calculations, we employ 22 single-electron states, which generate 231 Slater determinants forming the basis of the two-electron Hilbert space.

In the simplified 2D model, the system is reduced to an effective two-dimensional geometry by assuming that the image charge $e\rho'(x,y) \approx -e \rho(x, y)$ is located at a distance $d_{image}$ from the 2DEG positioned at $z=0$. In this case, the potential 
\begin{equation}
V_{i}(x,y)= \frac{-e}{4 \pi \varepsilon_{2DEG} \varepsilon _0} \int  \frac{\rho(x',y')dx' dy'}{\sqrt{(x-x')^2+(y-y')^2+d_\mathrm{image}^2}}
\label{eq:Vimage}
\end{equation}
with electron density
\begin{equation}
\rho(x,y)=2\int {dx_1dy_1} |\Psi(x_1,y_1,x,y)|^2,
\label{eq:rho}
\end{equation}
where $\Psi$ is the wave function of ground state of the two-electron problem $\hat{H}_2\Psi=E\Psi$.
Finally, the total energy
\begin{equation}
    E_t=E-\frac{e}{2} \int dxdy \rho(x,y) V_i(x,y).
    \label{eq:etotal}
\end{equation}

The system of Eqs.~(\ref{eq:Hamiltonian_k_space}), (\ref{eq:two_electrons}), and (\ref{eq:Vimage}) is solved self-consistently following the procedure described below. We start from the potential determined by the confinement $V=V_c$, compute the single-electron spectrum (\ref{eq:Hamiltonian_k_space}), and then solve the two-electron problem (\ref{eq:two_electrons}) to determine the ground state. From the resulting electron density, we evaluate $V_{i}$ (\ref{eq:Vimage}), which is then inserted back into the single-electron Hamiltonian (3). This cycle is repeated until convergence is reached. Finally, the total energy is determined using Eq.~(\ref{eq:etotal}). In calculations, we used the following material parameters, corresponding to the LAO/STO interface~\cite{Diez2015}: $t_h = 40$~meV, $t_l = 875$~meV, $t_d = 40$~meV, $\Delta_\text{E} = 47$~meV, $\Delta_\text{SO} = 10$~meV, $\Delta_\text{RSO} = 20$~meV.\\

\textbf{\textit{Results}} - Let us consider two electrons in the QD defined by the parabolic confinement potential, $V_c=\frac{1}{2}m\omega^2(x^2+y^2)$, where $m=0.286m_e$ is the effective mass associated with the $d_{xy}$ orbital. The corresponding characteristic length of the confinement is given by $\xi=\sqrt{\hbar / m \omega}$.

To quantify the negative charging energy, we describe the effective interaction with a single parameter $U = E_\text{t}^{2e} - 2E_\text{t}^{1e}$ corresponding to that used in the Anderson model~\cite{Jespersen2020,Guenevere2017,Cheng2015,Cheng2016}. Here, $2E_\mathrm{t}^{1e}$ is treated as the reference energy, denoting the ground-state energy of two separate one-electron QDs. This reference energy is crucial since, in our model, a single electron already polarizes  STO, inducing a self-interaction potential [Eq. ~(\ref{eq:Vimage})] through a self-consistently determined electron density $\rho_\text{1e}(x,y) = |\psi_\text{1e}(x,y)|^2$. 
When two electrons occupy the QD, their mutual Coulomb repulsion is counteracted by their joint polarization of the STO substrate. For a negative charging energy ($U<0$) to emerge, the two-electron image-charge interaction must be sufficiently strong to overcome the energetic cost associated with the direct Coulomb repulsion and the renormalization of the orbital energy due to the modified spatial profile of the two-electron wave function. Importantly, this energetic cost is partially offset by the one-electron image-charge contribution contained in the reference energy $2E_t^{1e}$, which lowers the energy of the single-electron states.
\begin{figure}
    \centering
    \includegraphics[width=1.0\linewidth]{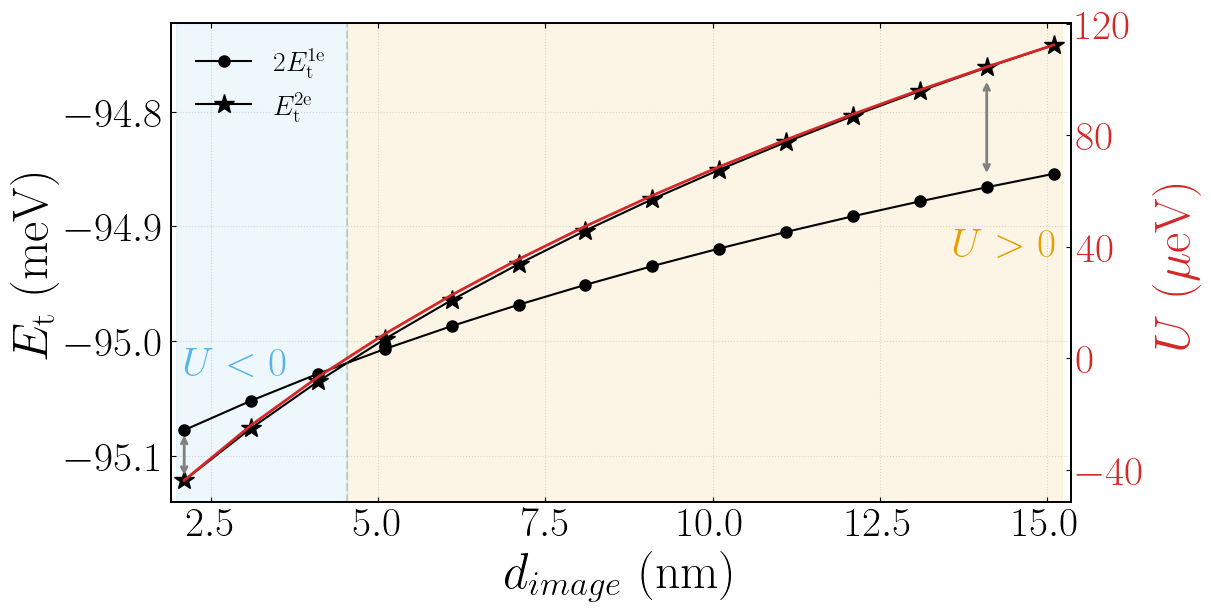}
    \caption{The total energy of two-electron QD (stars) and two single-electron QDs (dots) as a function of the distance of the image charge with respect to the 2DEG, $d_{image}$. Results for the QD with parabolic confinement defined by the characteristic length of $\xi=30$~nm. The blue-shaded region corresponds to the negative-$U$ regime, characterized by a preference for double occupancy; see the right-hand axis.}
    \label{fig2}
\end{figure}
\begin{figure}[!b]
    \centering
    \includegraphics[width=1.0\linewidth]{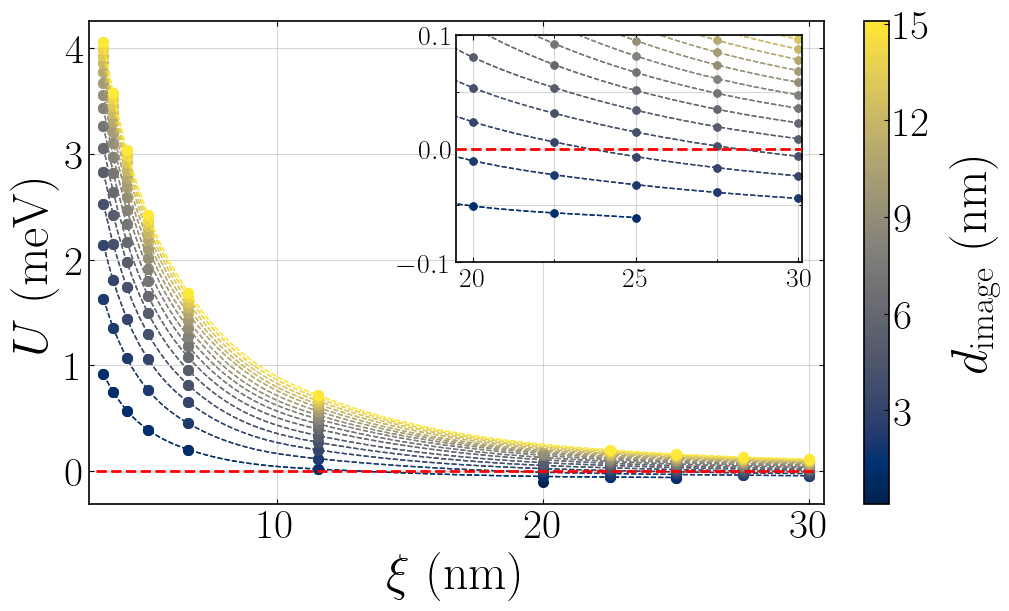}
    \caption{The interaction parameter $U=E_t^{2e}-2E_t^{1e}$ as a function of the characteristic confinement length $\xi$ for several selected values of $d_{\mathrm{image}}$, marked by color. The inset provides a magnified view of the region around the transition to the negative-$U$ regime. Dashed lines are guide to the eye.}
    \label{fig3}
\end{figure}

In Fig.~\ref{fig2}, we present the total energy of two-electron QD ($E_t^{2e}$) and two single-electron ($2E_t^{1e}$) QDs as a function of the image-charge distance, $d_{\mathrm{image}}$, for an exemplary confinement characterized by the length $\xi = 30\,\mathrm{nm}$. The results in Fig.~\ref{fig2} show that below a certain critical value of $d_{\mathrm{image}}$, the double occupancy of the dot lowers the total energy, indicating the emergence of the negative charging energy. This negative-$U$ regime is indicated by the blue-shaded region in Fig.~\ref{fig2}. Its extent is governed not only by the image-charge distance $d_{\mathrm{image}}$ but also by the characteristic confinement length $\xi$. In the limit of $\xi \rightarrow 0$, Coulomb repulsion dominates over attractive image-charge interaction as the latter remains finite due to the nonzero $d_\mathrm{image}$ in Eq.~(\ref{eq:Vimage}).

\begin{figure}[t]
    \centering
    \includegraphics[width=0.95\linewidth]{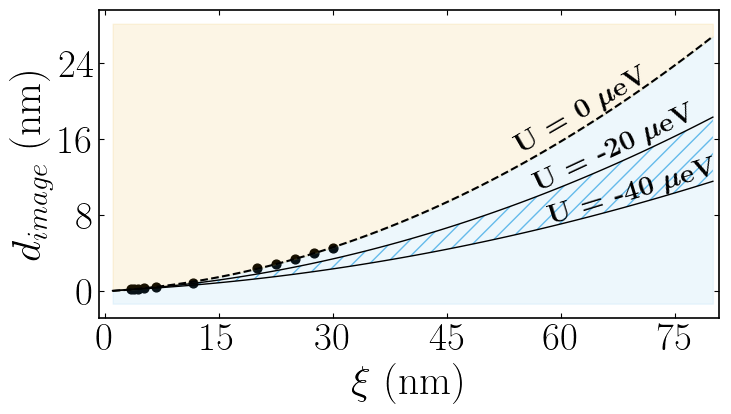}
    \caption{The critical value of $d_{\mathrm{image}}$ below which the negative-$U$ regime emerges  as a function of the characteristic confinement length $\xi$. The yellow and blue regions correspond to the positive and negative $U$ regimes, respectively.  The symbols correspond to numerical data, while the dashed curve represents an interpolation. The shaded region with $U \in [-40, -20]~\mathrm{\mu eV}$ indicates the range of the phase diagram extracted in the previous experiments.}
    \label{fig4}
\end{figure}

To analyze this dependence in detail, in Fig.~\ref{fig3} we present the parameter $U$ as a function of the characteristic confinement length $\xi$ for several selected values of $d_{\mathrm{image}}$, marked by color. We find that increasing the confinement length $\xi$ increases the critical value of $d_{\mathrm{image}}$ below which the negative-$U$ regime emerges. For each curve in Fig.~\ref{fig3}, we extract this critical value of $d_{\mathrm{image}}$ and summarize it in the form of a phase diagram presented in Fig.~\ref{fig4}, where the yellow and blue regions correspond to the positive and negative $U$ regimes, respectively. Due to the computational limitations, the calculations were performed only for a set of discrete values of $\xi$, shown as points in Fig.~\ref{fig4}. The dashed curve represents an interpolation that defines the boundary between the negative-$U$ and positive-$U$ regimes.

We note that, since our analysis is restricted to small and intermediate QD sizes, $\xi$, we approximate the extrapolation curve in this regime by a Taylor expansion up to second order. This approximation should therefore not be interpreted as implying that, for large $\xi$, double occupancy becomes favorable for arbitrarily large $d_{image}$. Instead, we hypothesize that the line $U=0$ reaches a maximum at some larger $\xi$, beyond which the charging energy increases and ultimately restores the instability of the doubly occupied state, $U>0$, in the absence of confinement, $\xi\rightarrow\infty$. This hypothesis is supported by two observations. First, $|U|$ does not become arbitrarily large as the QD size increases; rather, the charging energy is strictly bounded from below by the asymptotic limit of $U(\xi)$, as shown in Fig.~\ref{fig3}. Second, our supplementary calculations for two free electrons represented by Gaussian wave packets indicate that such a configuration is unstable, further supporting the conclusion that the apparent divergence of the $U=0$ extrapolation is not physically meaningful. Determining the position of this maximum, however, involves QDs substantially larger than those accessible within our present numerical approach and is therefore beyond our current computational capabilities. Therefore, the behavior of the transition between positive and negative charging energy in this large-QD regime remains an open question.\\

Finally, we emphasize that the characteristic lengths $\xi$ and $d_{\mathrm{image}}$ can be interpreted as the QD size and the thickness of the 2DEG, respectively. Both length scales are determined experimentally and have been reported in previous studies \cite{Sing2009, Salluzzo2009}. In particular, the latest paper on QDs in micro-membranes \cite{Meucci2026NegativeU} estimates the size of the QD in the range $20 - 80\,\mathrm{nm}$ and thickness of 2DEG  below $30~\mathrm{nm}$. 
Fitting the experimental spectrum of the QD within the Anderson model yields a negative-$U$ magnitude ranging from approximately $-40$ to $-20\,\mathrm{\mu eV}$~\cite{Meucci2026NegativeU}. The corresponding region of the phase diagram for these experimentally relevant parameters is marked in Fig.~\ref{fig4} by the dashed area. Notably, for a quantum dot size estimated in Ref.~\cite{Meucci2026NegativeU}, the experimentally relevant range of $U$ predicts a 2DEG thickness of approximately $5$–$15\,\mathrm{nm}$, which is in good agreement with realistic values. This analysis indicates that all quantities, including the characteristic lengths $\xi$ and $d_{\mathrm{image}}$, as well as the extracted value of $U$, remain within the experimentally realistic ranges, supporting the validity of the presented model.\\

Our findings raise the question of whether the negative-$U$ behavior observed in LAO/STO QDs represents a local manifestation of the pairing physics underlying superconductivity in oxide interfaces. Establishing such a connection requires understanding how the microscopic mechanism responsible for the negative charging energy relates to the mechanisms proposed for electron pairing in STO-based systems.

Up to date, several pairing mechanisms have been proposed for STO, including those mediated by optical phonons and quantum ferroelectric fluctuations~\cite{Edge2015}, as well as polaronic mechanisms arising from strong electron–phonon coupling~\cite{Devreese_2009}. Pairing driven by dynamic dielectric screening has also been proposed in the context of dilute superconductivity in STO~\cite{Takada1980}. Note, however, that the mechanism proposed here is purely electrostatic and does not rely on the exchange of dynamical bosonic excitations. The negative charging energy originates from the self-consistent interaction of electrons with image charge induced by the large dielectric response of STO, and is therefore controlled by the confinement geometry, and the finite thickness of the interfacial electron gas. Consequently, it cannot, in general, be reduced to a homogeneous dielectric-screening correction and provides an independent route to the negative-$U$ regime in oxide quantum dots. 

Finally, our results seem to be intriguing in light of recent proposals emphasizing electron-phonon coupling to polar optical modes as a potential origin of superconductivity in KTO-based interfaces~\cite{Liu2023,Zhou_2025}. In this context, an important open question is whether the image-charge interaction represents an independent mechanism, or rather the electrostatic manifestation of the same underlying polarization dynamics associated with optical phonons in STO. 
Establishing such a connection could provide a unified microscopic framework to understand both negative-$U$ behavior in QDs and superconducting pairing at oxide interfaces.\\

\textbf{\textit{Summary}} - In this paper, we propose a mechanism for the negative charging energy in LAO/STO QDs arising from image charges induced by the paraelectric properties of STO and the abrupt change in the dielectric constant across the interface. Within a configuration-interaction approach combined with a self-consistent treatment of the electrostatic environment, we show that the competition between direct Coulomb repulsion and the interaction with the image charge can drive a negative-$U$ regime. The emergence of this effect is controlled by two characteristic length scales: the QD confinement length, $\xi$, and the effective image-charge distance, $d_{\mathrm{image}}$, which is controlled by the thickness of the 2DEG.
The resulting phase diagram reveals a broad parameter regime in which the negative charging energy could appear, with experimentally reported values of $U$ ranging from $-40~\mathrm{\mu eV}$ to $-20~\mathrm{\mu eV}$. Importantly, for this range of $U$, the relevant length scales fall within the experimentally reported values for the LAO/STO QD size and the 2DEG thickness, suggesting that image-charge effects provide a reliable microscopic route to negative-$U$ behavior in LAO/STO QDs.

\section*{Acknowledgments}
This work is financed by the Horizon Europe EIC Pathfinder under the grant IQARO number 101115190 titled "Spin-orbitronic quantum bits in reconfigurable 2D-oxides". N.B. was supported by the ANR project SURIKAT (ANR-23-CE30-0036), ANR project NOBLESSE (ANR-24-CE92-0066) and the French-India CEFIPRA (project No: 7104-1). We gratefully acknowledge Poland’s high-performance computing infrastructure PLGrid (HPC Center Cyfronet) for providing computer facilities and support within computational grant no. PLG/2026/019641. We acknowledge that schematic illustration in Fig.~\ref{fig1}(b), apart from the labels, was generated using Google Deep Mind's Nano Banana 2.

\section*{Data Availability}
The data that support the findings of this article \cite{Negative_U_dataset} and the implementation of numerical scheme \cite{LAO_STO_QD_repo} are openly available.

%

\end{document}